\documentclass[a4page, 10pt]{article}
\usepackage{amssymb,amsmath}
\usepackage{graphicx}
\begin{document}

\title{Is Reality Digital or Analog?}

\author{Jarmo M\"akel\"a\footnote{Vaasa University of Applied Sciences,
Wolffintie 30, 65200 Vaasa, Finland, email: jarmo.makela@puv.fi}}

\maketitle

\begin{abstract}

A report of a discussion with Isaac Newton.

\end{abstract}



\bigskip

     Some years ago I happened to be listening to a lecture given by Gregory Chaitin on
what he called as "digital philosophy." \cite{yksi} In the following night, just before falling 
asleep, I was lying on my bed wondering what Isaac Newton, the discoverer of calculus and 
arguably the greatest physicist of all times, would have said about Chaitin's ideas.$^{1)}$ 
Gradually, my thoughts took a form of a firm decision: I would go and see Isaac Newton in person and ask 
from himself.

     It was not so easy to arrange an interview with Isaac Newton. He had a reputation of a
difficult person, and it was very uncertain, whether he would receive me in the first place. 
Then I got it: Newton would certainly be interested in the physics of our times. I decided to 
send him a collection of the best possible textbooks of physics I could find, together with 
copies of the original papers on black hole entropy and radiation written by Bekenstein and 
Hawking. \cite{kaksi, kolme} I also sent to Newton a letter, where I introduced myself and 
inquired for a possibility to meet him in person.
 
     A month later there was a letter in my mailbox. It read:

\bigskip

    \,\,\,\,\,\,"Will receive you with pleasure. Meet me in my London residence on the 18th
  of November,
      at 3 p. m. in 1700.

      \,\,\,\,\, Sincerely Yours,

\bigskip

       \,\,\,\, Isaac Newton."

\bigskip

I am not going to bore the reader with a detailed description of how I managed to arrive to London 
of the year 1700. To make a long story short I just tell that exactly at 3 p. m. on the 18th of 
November in 1700 I was standing on the doorsteps of Isaac Newton. With some hesitation I knocked 
the door...

  To my astonishment the door was opened by a beautiful lady in her early twenties. I introduced 
myself, and a warm, welcoming smile came to her face. "Please, come in," she said. "My uncle 
will receive you in his study. This way, please." At that point I remembered that 
some time after Newton had left from Cambridge to London to become a Warden and, later, a Master 
of Mint, his niece Catherine Barton moved into his house as a housekeeper. \cite{nelja} 
Together with Ms. Barton I ascended the stairs to Newton's study. "I am so happy that you sent 
all those books to my uncle", she said exuberantly. "My uncle has been somewhat depressed lately,
but ever since he received your books he has been in high spirits."

   We arrived to Newton's study. Ms. Barton knocked the door and we heard a gentle voice
"come in." She opened the door, I stepped in, and there was the great man himself. 

   From everything I had read I expected to see a short and somewhat aloof and inward bound 
person with an untidy appearance. On the contrary, he seemed extremely alert, he was 
well-dressed, and his long, white hair was carefully combed. He was very short, it is true, 
hardly five feet, but one immediately forgot his short stature after looking at his large, 
bright and intelligent eyes. 

   There was no need to formal introductions, because I had already introduced myself in a 
letter, and Isaac Newton certainly did not need any introductions. So I asked right away: "Is 
reality digital or analog?" Without any moments of hesitation he said: "Why, digital of course."
I was rather taken aback by his quick answer and asked: "How do you know?" "Because I have 
calculated it," was his response.

   This was quite a news. With some enthusiasm I asked, whether he would show me his 
calculation. "Of course," he said. In his study he had a large, white, wooden box. When he 
opened the box, I observed that it was full of papers. He perused the contents of the box for
a while and said: "No, it is not here. Now I remember: I performed the calculations on yesterday 
evening by the fireplace, and before going to sleep I left my calculations on the mantelpiece." 
He went to see the fireplace. "No, it is not here, either. Where can it be?" He opened the door and 
shouted: "Catherine!" "Yes, uncle," said Ms. Barton. "Have you seen the papers, which I left on 
the mantelpiece last night?" asked Newton. "Let me think," said Ms. Barton. "There was indeed 
some papers on the mantelpiece this morning. To me they seemed to be of no importance, and so I 
burned them in the fireplace."

   Isaac Newton went back to his study. He looked out of the window, and for a moment he 
seemed utterly crushed. Then, suddenly,
he burst into a laugh. I was so surprised that I could not avoid of beginning to laugh, too. We 
laughed together, and finally he sat on an armchair, gesturing me to do the same.

  "I became convinced of the digital nature of reality already in my youth," he said at last. "I
made some experiments with prisms and sunlight, you know, and I observed that although white light, 
when passing through a prism, dissolves into light with different colours, the colours thus produced 
cannot be dissolved further by another prism. For instance, red light stays red, while passing
through another prism, blue light stays bölue, and so on. So it was natural to conclude
that light consists of particles. Different particles produce different colours, and 
white light is produced as a mixture of those particles. As I am aware now, my chain of reasoning
was far from flawless, and Robert Hooke was actually quite correct in many points in his 
critique of my particle theory of light.$^{2)}$ Nevertheless, it took me into a whirl of 
speculations. 
If light consists of particles, then why should not matter as well? Maybe all matter is made of
some ultimate constituents with a very small number of intrinsic properties, I thought, and
all properties of matter may finally be reduced to the properties of its constituents.  In this 
sense, the behavior
of matter is digital, rather than analog. At that time I had no experimental support for my 
speculations, but reading the books you kindly sent me I learned that light consists of particles, 
which are known as photons, and all matter is made of atoms. So I was right after all, wasn't I?"
concluded Newton with a grin.  

  "Yes, you were, indeed," said I. "And what are your current thoughts about space and time?"

   "When I read about Einstein's special and general theories of relativity,"
he replied, "I was very impressed. From the very beginning I liked those theories enormously. I
had no difficulties whatsoever to accept them. Indeed, so simple and so 
beautiful those theories are that I wonder why I did not discover them by myself, when I was in
my prime of invention. The Equivalence Principle, for instance, was entirely within my grasp.
Unfortunately, the idea of the constancy of the speed of light with respect to all inertial
observers never dawned to me. but even that I could have deduced, if I had had the courage to
draw the ultimate conclusions from my first law of motion implying, in effect, an equivalence
of all inertial observers."$^{3)}$ 

     As a specialist in general relativity with an avid interest in the history of science I 
was forced to admit to myself that Newton was not bragging, but simply telling the truth. He 
thought
for a moment, and then he continued:

   "The ultimate problem, of course, is how to make general relativity compatible with quantum
mechanics.$^{4)}$ To me it seems likely that spacetime, in the same way as matter, has a sort of atomic
structure. In other words, there must be some ultimate constituents of space and time."

      "That is very interesting," said I. "Why do you think so?" 

      "When speculating on the possible atomic structure of spacetime," he said, "one is very
much in the same position as I was in my youth when speculating on the possible atomic structure
of matter. There is no direct observational evidence. However, there is a very important piece
of indirect theoretical evidence."

       "And what is that?" asked I.

        "The result, obtained by Stephen Hawking, my worthy successor in the Lucasian chair,
that black holes emit spontaneous thermal radiation. I must confess that understanding Hawking's
important result was very hard to me. However, I think that I have now mastrered every detail
of its standard derivation, \cite{kolme} and I am convinced that it is true. Black hole indeed has certain
temperature, which is inversely proportional to its mass. And it has entropy, too, which - in 
the natural units, where all natural constants are set to unity - is one-quarter of its event 
horizon area."$^{5)}$

         "And what does that have to do with the atomic structure of space and time?" asked I.

         "Do you still remember the thermodynamical reason for the particle nature of light?" 
asked Newton.

         "Yes," said I. "When considering the thermodynamical properties of electromagnetic
radiation - light, for example - one finds that if one adds up the energies carried by the 
different frequencies of the radiation in a given temperature, the total energy density of the
radiation becomes infinite, which is absurd, of course. However, if one assumes that the 
radiation consists of particles, each of them carrying an energy, which is proportional
to the frequency of the radiation, the energy density miraculously becomes finite, and the 
resulting expression for the intensity of the radiation agrees with observations." 
            
      "Precisely," said Newton. "When considering the thermodynamical properties of black holes,
one meets with similar unphysical infinities." At this point he went to his white wooden
box and pulled out an empty sheet of paper. From his desk he picked up a quill pen, beginning
to write equations on the paper. "The 
thermodynamical properties of any system," he continued, "may be deduced from its partition
function
\begin{equation}
Z(\beta) = \sum_E g(E)e^{-\beta E},
\end{equation}
where $\beta$ is the inverse temperature of the system, and we have summed over all possible
energies $E$ of the system.$^{6)}$ Of course, if the energy spectrum is continuous, the sum must be
replaced by an integral. $g(E)$ tells the degeneracy of a state with energy $E$. In other
words, it tells the number of the microscopic states associated with the same total energy 
$E$ of the system. Can you recall what is entropy?"

    "Well," said I, "if all microstates associated with the same macroscopic state of the system
have equal probabilities, then the entropy of the system is, in natural units, the natural 
logarithm of the number of those microstates..."

     "...which means that the number of the associated microstates is the exponential of the
entropy," said Newton. "As you may remember, the radius of the Schwarzschild black hole, which
is the simplest possible black hole is, in natural units, $R = 2M$, where $M$ is the mass 
of the hole.$^{7)}$
Hence it follows that the event horizon area of the hole is $A = 4\pi R^2 = 16\pi M^2$. Because
the entropy of a black hole is one-quarter of its event horizon area we find, identifying the 
mass $M$ of the hole with its total energy $E$, that the partition function of the Schwarzschild
black hole is
\begin{equation}
Z(\beta) = \sum_E e^{4\pi E^2} e^{-\beta E}.
\end{equation}
As far as we know, there is no upper limit for the mass of a black hole, and so we observe that
the sum in our partition function diverges.$^{8)}$ Actually, it diverges very badly, because the function
$e^{4\pi E^2}e^{-\beta E}$ goes very rapidly towards the positive infinity, when the energy $E$
is increased, no matter what is $\beta$."

      Newton laid his quill pen on his writing desk and concluded: "Since the sum in Eq. (2) 
diverges, a black hole has no well-defined partition function. Without partition function we
cannot deduce its thermodynamical properties. Hence we meet with an unphysical infinity, which
is somewhat similar to the one we encounter when considering the energy density of 
electromagnetic radiation."

      "Do you have any ideas of how to get rid of that infinity?" asked I.

      "The reason for a divergent partition function," replied Newton, "was that we identified
$g(E)$ with the exponential of the black hole entropy, which was assumed to be one-quarter of its
event horizon area. One is therefore inclined to doubt the general validity of the law of the 
simple proportionality between the area and the entropy of a black hole. Of course, it is valid
when the temperature of the hole is very low. There is no question about that. An interesting 
issue is, whether the area law of black hole entropy holds even when the temperature of the hole is
very high. In my opinion we should modify the function $g(E)$ such that the resulting entropy
of the black hole is proportional to the event horizon area at low temperatures, but not 
necessarily at high temperatures. The main goal is to make the partition function convergent."

       "How are you going to find an appropriate modification of $g(E)$?" asked I.

        "I construct the event horizon out of discrete constituents," answered Newton, picking up
his quill pen again, and moisturing it with ink. "Each constituent is assumed to carry an area
which, in the natural units, is an integer times an appropriate constant, and the total area
of the horizon is assumed to be the sum of the areas of those constituents. In other words, we
write the total area of the horizon as:
\begin{equation}
A = 16\pi\alpha^2 (n_1 + n_2 + n_3 + ... + n_N),
\end{equation}
where $N$ is the number of the constituents, and $\alpha$ is a pure number to be determined 
later. The quantum numbers $n_1, n_2, ..., n_N$ are non-negative integers, and they determine 
the areas of the individual constituents of the horizon. More precisely, if we pick up a 
constituent $j$, $(j = 1, 2, ..., N)$ the area contributed by that constituent to the total area
of the horizon is $A_j = 16\pi\alpha^2 n_j$, where $n_j$ is a non-negative integer."$^{9)}$

      "To me your idea seems very simple and natural," said I. "What is now the degeneracy of 
those states of the black hole, which have the same energy?"

       "Because the event horizon area of a black hole is $A = 16\pi M^2$," replied Newton, "the
mass $M$ of the hole is determined by its horizon area. Identifying the mass of the hole with
its energy we find, using Eq. (3), that the possible energies of the hole are 
\begin{equation}
E_n = \alpha\sqrt{n},
\end{equation}
where $n$ is a non-negative integer determined by the quantum numbers $n_j$ such that
\begin{equation}
n = n_1 + n_2 + n_3 + ... + n_N.
\end{equation}
In other words, the energy of the hole is uniquely determined by the sum of the quantum numbers
$n_j$. There are several ways to write the non-negative integer $n$ as a sum of $N$ non-negative
integers $n_j$, and this gives a rise to the degenerate states."

     Newton moistured his quill pen again with ink and then continued: "When a constituent is
in vacuum, {\it i. e.} $n_j = 0$, it does not contribute to the physical properties of the 
horizon, and hence it is natural to identify the microscopic states with the different 
combinations of the non-vacuum states of its constituents. As a consequence, the degeneracy
$g(E_n)$ of a state with energy $E_n$ is the number of ways of writing the positive integer
$n$ as a sum of at most $N$ positive integers $n_j$. More precisely, it is the number of ordered
strings $(n_1, n_2, ..., n_m)$, where $n_1, n_2, ..., n_m$ are positive integers, 
$1 \leq m\leq N$, and $n_1 + n_2 + ... + n_m = n$. The number of ways of writing a positive
integer $n$ as a sum of $m$ positive integers is the same as is the number of ways of arranging
$n$ balls in a row in $m$ groups by putting $(m-1)$ divisions in the $(n-1)$ empty spaces between the
balls. The position for the first division may be chosen in $(n-1)$ ways, for the second in
$(n-2)$ ways, and so on. The total number of the combinations for the positions of the 
divisions is therefore
$(n-1)(n-2)\cdots(n-m+1)$. However, because the divisions are identical, we must divide this number by the
number of the possible orderings of the divisions, which is $(m-1)(m-2)\cdots2\cdot 1 = (m-1)!$. 
Hence the total number of ways of writing a positive integer $n$ as a sum of exactly $m$ positive 
integers is$^{10)}$
\begin{equation}
 \left(\begin{array}{cc}n-1\\m-1\end{array}\right) := \frac{(n-1)(n-2)\cdots(n-m+1)}{(m-1)!},
\end{equation}  
and the degeneracy of the state with energy $E_n$ is
\begin{equation}
g_1(E_n) := \sum_{m=1}^N\left(\begin{array}{cc}n-1\\m-1\end{array}\right),
\end{equation} 
provided that $N\leq n$. The number $m$ cannot be greater than $n$, and therefore the degeneracy
is
\begin{equation}
g_2(E_n) := \sum_{m=1}^n\left(\begin{array}{cc}n-1\\m-1\end{array}\right),
\end{equation}
whenever $N\geq n$. So we obtain an expression
\begin{equation}
Z(\beta) = \sum_{n=1}^N\left\lbrack \sum_{m=1}^n\left(\begin{array}{cc}n-1\\m-1\end{array}\right) 
e^{-\beta\alpha\sqrt{n}}\right\rbrack + \sum_{n=N+1}^\infty\left\lbrack\sum_{m=1}^N
\left(\begin{array}{cc}n-1\\m-1\end{array}\right)e^{-\beta\alpha\sqrt{n}}\right\rbrack
\end{equation}
for the partition function of a black hole."

   "And your partition function is convergent?" asked I.

    "Absolutely," said Newton. "Using Eq. (6) we find that when $n\gg N$, we have$^{11)}$
\begin{equation}
\sum_{m=1}^N\left(\begin{array}{cc}n-1\\m-1\end{array}\right) \sim \frac{1}{(N-1)!}(n-1)^{N-1}.
\end{equation}
For fixed $N$ the sum
\[
\sum_{n=N+1}^\infty (n-1)^{N-1}e^{-\beta\alpha\sqrt{n}}
\]
will certainly converge, and therefore our partition function will converge as well."

   "So you have managed to show that if the event horizon of a black hole consists of a finite
number of discrete constituents, then the partition function of the hole is well-defined and 
finite," said I. "Very fine. What about the entropy of the hole at low temperatures?"

    "You are familiar with the Binomial Theorem, I suppose?" asked Newton.

     "Of course," said I. "If $n$ is a positive integer, the Binomial Theorem implies that
\begin{equation}
(a+b)^n = \sum_{k=0}^n\left(\begin{array}{cc}n\\k\end{array}\right)a^{n-k}b^k
\end{equation}
for any reals $a$ and $b$."

      "When the temperature of the black hole is very low, we have $\frac{n}{N}\ll 1$, which 
means that a great majority of the constituents of the event horizon are in vacuum," said Newton.
"In this limit the degeneracy of a state with energy $E_n$ is given by the function $g_2(E_n)$
of Eq. (8). Putting $a = b = 1$ in the Binomial Theorem we find:
\begin{equation}
g_2(E_n) = \sum_{m=1}^n\left(\begin{array}{cc}n-1\\m-1\end{array}\right) = (1 + 1)^{n-1} 
= 2^{n-1},
\end{equation}
and hence the entropy of the hole is, in the low temperature limit:
\begin{equation}
S = \ln(2^{n-1}) = (n-1)\ln 2.
\end{equation}
Since the event horizon area $A = 16\pi\alpha^2 n$, we get for large $n$, taking $\alpha = 
\frac{1}{2}\sqrt{\frac{\ln 2}{\pi}}$:
\begin{equation}
S = \frac{1}{4}A,
\end{equation}
which is the well-known law for black hole entropy."

   "Excellent!" I exclaimed. "That is really impressive!"

    "Elementary," said Newton. "Anyway, our calculation may be used as an argument for an 
atomic structure of spacetime. Assuming that the event horizon 
of a black hole consists of a finite number of discrete constituents, we may both remove the
unphysical infinities from its partition function, and obtain a correct expression for the black
hole entropy at low temperatures. If the event horizon of a black hole consists of some 
fundamental constituents, then why should not spacetime as a whole as well? After all, effects
very similar to the spontaneous thermal radiation of black holes may be observed, at least in 
principle, even in flat spacetime, where no black holes are present. When reading Wald 
\cite{yksitoista} I learned
that an observer in a uniformly accelerating motion will detect thermal radiation of particles
even when all inertial observers detect a vacuum.$^{13)}$ To me it seems likely that to 
explain effects like this one must assume a sort of atomic structure of spacetime. In this sense 
spacetime, in the same way as matter and indeed reality itself, is digital, rather than analog."

  "Although I am very impressed by your derivation of the area law of black hole entropy at 
low temperatures, I must confess that I am still sceptical," said I. "In Einstein's general 
relativity the concept of distance plays a fundamental role in the sense that if we know the 
distances between the points of spacetime, we know the so-called metric tensor $g_{\mu\nu}$
of spacetime. The metric tensor, in turn, determines the geometric and the causal properties 
of spacetime.$^{14)}$ If spacetime really has an atomic structure, then
what happens to the concepts of distance, time and causality?"

  "When the appropriate natural constants are put in the calculations we just made," said Newton,
"one finds that the possible areas of the constituents of the black hole event hoizon may be 
expressed in terms of the quantum numbers $n_j$ as $4n_j(\ln 2)\ell_{Pl}^2$, where the quantity
$\ell_{Pl}$ is written in terms of the gravitational constant $G$, Planck's constant $\hbar$ and 
the speed of light $c$ in the form:
\begin{equation}
\ell_{Pl} := \sqrt{\frac{\hbar G}{c^3}} \approx 1.6 \times 10^{-35}m.
\end{equation}
As I have read from your books, this quantity is known as the Planck length. The result suggests 
that at the Planck length scale the concepts of distance, time and causality will probably 
lose their meaning. However, at macroscopic scales, where the number of the constituents of the 
spacetime region under consideration is very large, these concepts may be used to describe the 
statistical properties of spacetime."

  "What will replace distance as the fundamental concept in your model?" asked I.

   "The concept of area," was Newton's firm answer. "At macroscopic scales one may reduce 
the concept of distance to the concept of area. Actually, that is a trivial consequence of the
fact that the dimension of spacetime at macroscpic scales is four."

    "To me that seems highly non-trivial," said I.

     "Well," said Newton, showing admirable patience to my slow intellect. "Consider first a 
tetrahedron. A tetrahedron is a three-dimensional object with 4 vertices not lying on the same 
plane. Each vertex is connected to every other vertex by an edge. As a result there are 6 edges 
and 4 triagles in a tetrahedron. A natural four-dimensional generalization of a tetrahedron
has 5 vertices not lying in the same three-dimensional space, each vertex being connected, again,
to every other vertex by an edge. As I read from {\it Gravitation}, \cite{viisitoista} this 
object is known as a 
four-simplex. An interesting property of a four-simplex is that the number of its edges is 10,
which is the same as is the number of its triangles.$^{15)}$ So there is a one-to-one 
relationship
between the edges and the triangles of a four-simplex, and we may not only express the triangle
areas in terms of the edge lengths, but the edge lengths may also be expressed in terms of the 
triangle areas. Hence we observe that whenever we pick up 5 points in a four-dimensional spacetime
such that those points do not lie in the same three-dimensional space, the distances between
the points may be expressed in terms of the triangle areas of a four-simplex having those 
points as its vertices. In this sense the concept of distance may really be reduced, in
four-dimensional spacetime, to the concept of area."

   "I see," said I. "It is trivial. I suppose it is natural to assume that the triangles in
your four-simplex, in the same way as the event horizons of black holes, consist of discrete 
constituents, each of them contributing an area of the form $4n_j(\ln 2)\ell_{Pl}^2$ to the 
triangle area?"

    "That is right,"said Newton. "As a consequence, we may reduce the distances between the points
of spacetime, and thereby the metric tensor $g_{\mu\nu}$, to the quantum states of those 
constituents. Since the metric tensor $g_{\mu\nu}$ determines the causal properties of spacetime,
it seems to me likely that there exists a still unknown law of nature, closely related to the 
Second Law of Thermodynamics, which implies that when we consider spacetime at macroscopic scales, 
the quantum states of its constituents are distributed in such a way that the everyday notions of 
time and causality are recovered. However, that is just speculation." 

    Obviously, our discussion was coming to an end. I got up from the armchair to thank Newton for 
an interesting discussion, when suddenly Ms. Barton rushed into the room. "Uncle," she said. "I 
managed to find your papers. They were not the ones, which I burned in the fireplace this 
morning, but you had left them on the breakfast table. Here they are!"

    To my surprise Newton handed the papers to me and said: "Please, keep these. I think that you 
will need them more than I. After all, it would be a strong violation of causality, if I 
published them by myself."

   "I think that you will find your way out by yourself," said Ms. Barton, observing that I was
just about to leave. "I will," said I. Saying good-bye to Isaac Newton and his charming niece I
left Newton's study and closed the door behind me.

   When descending the stairs I could not resist the temptation of having a look at those
papers. Obviously, they constituted a finished draft of an extensive research report, written 
with a miniature, but still clear and precise hand. Its front page carried the title:

\bigskip
  
THE COMPLETE QUANTUM THEORY OF GRAVITATION

\bigskip

Near the end of the draft my eyes picked up a sentence: "So we observe that the fundamental 
equation of quantum gravity is..."

   I stumbled. I fell on my face and came rolling down the stairs. I felt pain in my shoulder 
and all papers flew out of my hands...

   I was lying on the floor beside of my bed. It was eight o'clock in the morning. 
          
\vfill

\eject

{\bf NOTES}

\bigskip

1) Isaac Newton lived during the years 1642-1727.

2) Robert Hooke (1633-1707) was a rival and an antagonist of Isaac Newton. His original 
achievements included, among other things, significant improvements on a microscope, and
a discovery of cells.

3) The Equivalence Principle states, in very broad terms, that it is not possible to decide,
on grounds of local observations, whether one is in a gravitational field, or in an accelerating
frame of reference. Historically, it was the starting point of Einstein's general theory of
relativity, which explains gravity by means of the properties of space and time. 
The constancy of the speed of light with respect to all inertial observers, in turn,
is the starting point of Einstein's special theory of relativity. Among other things, it implies
that space and time are not absolute, but they depend on the observer's state of motion. This
is in marked contrast with Newton's classical mechanics, where space and time are considered
absolute. Hence Newton's favorable opinion on Einstein's special and general theories of 
relativity is most remarkable.

4) A theory, which makes general relativity and quantum mechanics compatible with each other
is known as a quantum theory of gravitation, or quantum gravity, in short. Construction of the
quantum theory of gravity is generally considered as the greatest challence of theoretical 
physics.
 
5) By definition, a black hole is a region of space, where the gravitational field is so 
strong that not even light can escape from that region. The boundary of a black hole is known
as event horizon. A black hole is created, among other things, if a very large star collapses
under its own weight after burning off its nuclear fuel. It was a great discovery of Stephen
Hawking that black holes are not completely black after all, but they emit radiation by means
of certain quantum-mechanical processes taking place in the vicinity of the event horizon. An 
importance of this result lies in the fact that it brings together general relativity, quantum
mechanics and thermodynamics, the fundamental theories of physics.

6) The parameter $\beta$ may be written in terms of the absolute temperature $T$ of a system, 
in natural units, as $\beta = \frac{1}{T}$. If we know the partition function of a system, we 
may obtain expressions for various thermodynamical quantities as functions of $T$. For instance,
the average energy of a system is
\[
E_{ave} = -\frac{d}{d\beta}\ln Z(\beta).
\]

7) In terms of the gravitational constant $G$ and the speed of light $c$ the radius of the 
Schwarzschild black hole may be written as
\[
R = \frac{2GM}{c^2}.
\]
For instance, if the mass $M$ of the hole is one solar mass, then $R$ is about three kilometers.

8) Solutions to the problem of diverging partition function  have been suggested in Refs. 
\cite{viisi} and \cite{kuusi}.

9) One of the consequences of Newton's model is that the event horizon area of a black hole has
an equally spaced spectrum. The idea of an equally spaced horizon area spectrum was raised first
by Bekenstein in Ref. \cite{seitseman}, and later by Bekenstein and Mukhanov in Ref. 
\cite{kahdeksan}.

10) For instance, the number 5 may be written as a sum of 3 positive integers in
\[
\left(\begin{array}{cc}5-1\\3-1\end{array}\right) = \frac{(5-1)(5-2)}{(3-1)!} = 6
\]
ways. Indeed, we have:
\[
5 = 2+2+1 = 2+1+2 = 1+2+2 = 1+1+3 = 1+3+1 = 3+1+1.
\]
Finding the microscopic states associated with the given macroscopic state of a black hole is
one of the greatest problems of black hole physics. The problem has been considered in the context
of string theory in Ref. \cite{kymmenen}, and in that of loop  quantum gravity in 
Ref. \cite{yksitoista}. It should be noted that in Newton's model different orderings of the same
quantum numbers $n_j$ represent different microscopic states.

11) When $n\gg N$, the leading term in the sum on the left hand side of Eq. (10) is the one,
where $m = N$. Using Eq. (6) one finds that the leading term of 
$\left(\begin{array}{cc}n-1\\N-1\end{array}\right)$ is given by the right hand side of Eq.(10). 

12) This is known as the {\it Unruh effect}. \cite{kaksitoista} An accelerating observer 
experiences the so-called Rindler horizon, which has properties very similar to those of the
event horizon of a black hole. It is possible that the Unruh effect could be explained by means
of an appropriate discrete model of spacetime. \cite{kolmetoista, neljatoista}

13) In general relativity one defines between spacetime points $(x^0,x^1,x^2,x^3)$ and 
$(x^0+dx^0,x^1+dx^1,x^2+dx^2,x^3+dx^3)$ the so-called {\it line element}
\[
ds^2 = \sum_{\mu,\nu = 0}^3 g_{\mu\nu}\,dx^\mu\,dx^\nu,
\]
where $g_{\mu\nu}$ is the metric tensor at the point $(x^0,x^1,x^2,x^3)$. The distance between
these points is the square root of the modulus of $ds^2$. A curve connecting two points of spacetime is 
{\it spacelike}, if the line element at its every point is positive, and {\it timelike}, if it 
is negative. Hence the metric tensor determines both the metric-, and the causal properties of 
spacetime. 

14) The numbers of edges and triangles of a four-simplex, respectively, are the same as are the
numbers of ways of picking up two and three vertices out of five. Two vertices may be picked up 
in $\frac{5\cdot 4}{1\cdot 2} = 10$ ways, and three vertices may also be picked up in 10  ways, 
because $\frac{5\cdot 4\cdot 3}{1\cdot 2\cdot 3} = 10$. 

\vfill

\eject

\end{document}